\documentstyle[preprint,prb,aps]{revtex}

\begin{document}
\draft
\title{Understanding light quanta:\\
First quantization of the free electromagnetic field}
\author{A. C. de la Torre}
\address{Departamento de F\'{\i}sica,
 Universidad Nacional de Mar del Plata\\
 Funes 3350, 7600 Mar del Plata, Argentina\\
dltorre@mdp.edu.ar}
\maketitle
\begin{abstract}
The quantization of the electromagnetic field in vacuum is
presented without reference to lagrangean quantum field theory.
The equal time commutators of the fields are calculated from
basic principles. A physical discussion of the commutators
suggest that the electromagnetic fields are macroscopic emergent
properties of more fundamental physical system: the photons.
\end{abstract}

\section{INTRODUCTION}
In a letter to his friend Besso in 1951, Einstein complained that
after fifty years of conscious meditation he had not come any
closer to the answer to the question \emph{What are light
quanta?} The level of understanding of quantum electrodynamics
reached at that time by the physics community was not sufficient
for him since he added: ``today any brute believes that he knows
the answer, but he is wrong''\cite{Ei}. In this work and in the
following ones we will present the understanding that we have
today on the quantum of electromagnetic radiation. This
understanding, perhaps, would still be unsatisfactory for
Einstein but at least we can claim that we understand the light
quanta as much, or as little, as we understand any other material
particle. The main idea in this search for clarity will be to
recognize that the electromagnetic fields are nor a primary
physical entity but are a collective manifestation of more
fundamental entities, the photons, carriers of energy, momentum
and angular momentum.

It is remarkable that in almost all quantum mechanics texts, even
in the advanced ones, the quantization of the electromagnetic
fields is avoided and the subject is postponed to lagrangean
quantum field theory. In few cases\cite{schi} an introduction to
quantum field theory is given just in order to present the
(second) quantization of the electromagnetic field. After having
learned about the success and power of quantum mechanics in the
description of physical systems, one should wonder whether the
same procedures can be successfully applied to the
electromagnetic fields. A didactic relevance of this work is to
show that it is possible, and easy, to quantize the free
electromagnetic field without any reference to quantum field
theory which is, indeed, a difficult subject. However, after
seing that the quantization of the electromagnetic field can, in
principle, be performed following the same line of thoughts as,
say, the harmonic oscillator, we will give arguments indicating
that this is not a meaningful thing to do because there are good
reasons for  thinking that the electromagnetic field is not a
primary physical system, that necessarily requires a quantum
treatment, but should be understood as a collective manifestation
of an ensemble of more fundamental quantum mechanical objects:
the photons.

In this work we will consider only observable quantities like the
electric and magnetic fields and we will avoid the use of
unobservable potential fields that present considerable
complications in a Lorentz invariant gauge fixing adequate for
quantization.
\section{THE CLASSICAL FREE ELECTROMAGNETIC FIELD}
For the description of the main features of the classical free
electromagnetic field we adopt Heaviside system, that is,
rationalized Gau\ss\ units\cite{units}. The six observables of
the physical system are the electric field vector ${\mathbf
E}({\mathbf r},t)$ and the magnetic field (pseudo)vector
${\mathbf B}({\mathbf r},t)$ in vacuum. These fields satisfy
Maxwell's equations
\begin{eqnarray}\label{mx1}
  -\varepsilon_{jkl} \partial_{k}E_{l}&=& \frac{1}{c}\partial_{t}B_{j}\ , \\
  \varepsilon_{jkl} \partial_{k}B_{l}&=& \frac{1}{c}\partial_{t}E_{j}\ , \\
 \partial_{k}E_{k}&=&0\ ,\\
\partial_{k}B_{k}&=&0 \ ,
\end{eqnarray}
where $\varepsilon_{jkl}$ is the well known total antisymmetric
(pseudo)tensor, $\partial_{k}= \frac{\partial}{\partial x_{k}}$,
$\partial_{t}= \frac{\partial}{\partial t}$ and we adopt
Einstein's convention, stating that repeated indices indicate a
summation. Furthermore, in order to keep the notation as simple
as possible we will suppress the space-time arguments of the
fields wherever they are not explicitly needed. At any point in
space where the fields don't vanish there is an energy and
momentum \emph{density} given by the local value of the fields;
 the \emph{total} energy $H$ and momentum ${\mathbf P}$ are given by
the volume integrals
\begin{eqnarray}
  H&=& \frac{1}{8\pi}\int\!\!\! d^{3}{\mathbf r}\
({\mathbf E^{2}}+{\mathbf B^{2}}) \\
  {\mathbf P} &=& \frac{1}{8\pi c}\int\!\!\! d^{3}{\mathbf r}\
({\mathbf E}\times{\mathbf B} - {\mathbf B}\times{\mathbf E}) \ .
\end{eqnarray}
Note that, although at the classical level it is unnecessary, in
the definition of momentum ${\mathbf P}$ we allow for a possible
noncommutation of the fields. In similar manner, we could define
the centroid of the energy distribution and the angular momentum
carried by the fields.

In our treatment of electromagnetism we will make intensive use
of the transformations of space inversion $\mathcal{P}$, time
inversion $\mathcal{T}$, charge conjugation $\mathcal{C}$ and the
\emph{dual} transformation $\mathcal{D}$ (defined below). The
fields have following transformation properties.

\begin{eqnarray}
 {\mathcal P}\left[ {\mathbf E}({\mathbf r},t)\right]&=&
{-\mathbf E}({-\mathbf r},t) \ ,\\
 {\mathcal T}\left[ {\mathbf E}({\mathbf r},t)\right]&=&
{\mathbf E}({\mathbf r},-t)\ ,\\
{\mathcal C}\left[ {\mathbf E}({\mathbf r},t)\right]&=&
{-\mathbf E}({\mathbf r},t)\ ,\\
{\mathcal D}\left[ {\mathbf E}({\mathbf r},t)\right]&=& -{\mathbf
B}({\mathbf r},t)\ ,
\end{eqnarray}
\begin{eqnarray}
  {\mathcal P}\left[ {\mathbf B}({\mathbf r},t)\right]&=&
{\mathbf B}({-\mathbf r},t)\ ,\\
 {\mathcal T}\left[ {\mathbf B}({\mathbf r},t)\right] &=&
{-\mathbf B}({\mathbf r},-t)\ ,\\
{\mathcal C}\left[ {\mathbf B}({\mathbf r},t)\right]&=&
{-\mathbf B}({\mathbf r},t)\ ,\\
{\mathcal D}\left[ {\mathbf B}({\mathbf r},t)\right]&=& {\mathbf
E}({\mathbf r},t)\ .
\end{eqnarray}
Since the set of (vacuum) Maxwell's equations is invariant under
these transformations, the transformation of any true relation
among the fields results in another true relation.

It is very convenient to introduce a complex electromagnetic
field ${\mathbf F}({\mathbf r},t)$ defined by
\begin{equation}\label{F}
  {\mathbf F}({\mathbf r},t)={\mathbf E}({\mathbf r},t)+
i{\mathbf B}({\mathbf r},t)\ .
\end{equation}
It has been claimed\cite{hert} that this is not only a very
convenient notation but it is a manifestation of a deep
mathematical structure, called Geometric Algebra, adequate for
modelling physical reality. Note that the real part of ${\mathbf
F}$ is a vector and the imaginary part is a pseudovector and that
the relation above can be trivially inverted in order to recover
the electric and magnetic field that may be more familiar to the
reader. For the components of the free electromagnetic field
$F_{k}$, Maxwell's equations are
\begin{eqnarray}\label{mx2}
  \varepsilon_{jkl} \partial_{k}F_{l}&=& i\frac{1}{c}\partial_{t}F_{j}\ ,  \\
 \partial_{k}F_{k}&=&0\ ,
\end{eqnarray}
 and the transformation properties are
\begin{eqnarray}
  {\mathcal P}\left[ {\mathbf F}({\mathbf r},t)\right]&=&
{-\mathbf F}^{\ast}({-\mathbf r},t)\ ,\\
 {\mathcal T}\left[ {\mathbf F}({\mathbf r},t)\right] &=&
{\mathbf F}^{\ast}({\mathbf r},-t) \mbox{ (classical)}\ ,\\
{\mathcal C}\left[ {\mathbf F}({\mathbf r},t)\right]&=&
{-\mathbf F}({\mathbf r},t)\ ,\\
{\mathcal D}\left[ {\mathbf F}({\mathbf r},t)\right]&=& {i\mathbf
F}({\mathbf r},t)\ .
\end{eqnarray}
The reason for emphasizing that the behaviour of ${\mathbf F}$
under time inversion is classic will be clarified later.
\section{FROM CLASSICAL OBSERVABLES TO OPERATORS}
We can now follow the standard procedure for the quantum
mechanical treatment of a physical system, but applied to the
free electromagnetic field. The first step is to define a set of
hermitian \emph{field operators} ${\mathbf E}({\mathbf r},t)$ and
${\mathbf B}({\mathbf r},t)$ in a Hilbert space, corresponding to
the electric and magnetic field observables (we use the same
symbols to denote the operators and the classical variables but
this should cause no confusion). In the Heisenberg picture, the
infinite set of operators is parameterized by the variables
$({\mathbf r},t)$ that are not dynamical variables but are
coordinates that characterize a space-time point. Therefore at
every space-time point $({\mathbf r},t)$  we have six operators
${\mathbf E}$ and ${\mathbf B}$ and we must find the commutation
relations for this infinite set. The commutation relations should
determine the eigenvalue spectrum related to the observations and
the eigenvectors related to the corresponding states of the
system.

In the derivation of the commutation relations, we will assume
that all relations and symmetries of the classical
electromagnetic fields are also satisfied by the corresponding
quantum operators. Every constraint on the classical fields
should also be present in the quantum description of the fields.
The opposite is not true. There are restrictions imposed in the
quantum treatment of a system that do nor appear in a classical
modelling. Quantum mechanics has stronger constrains and
correlations that have manifestation in characteristic quantum
effects like nonseparability, contextuality, uncertainty
relations among classically unrelated quantities, etc. Clearly,
the imposition of all classical relations and symmetries is not
sufficient to determine the commutation relations. If it where
so, in  sense, quantum mechanics could be derived from classical
mechanics; therefore somewhere we must introduce a quantization
postulate that in our case will appear as a mathematical
\emph{ansatz} concerning the form of the commutators.

In terms of the field ${\mathbf F}$, the hamiltonian and the
momentum operators are given by
\begin{eqnarray}
  H&=& \frac{1}{16\pi}\int\!\!\! d^{3}{\mathbf r}\
  ({\mathbf F^{\dag}}\cdot{\mathbf F}+{\mathbf F}\cdot{\mathbf F^{\dag}})\ ,\\
  {\mathbf P} &=& \frac{-i}{16\pi c}\int\!\!\! d^{3}{\mathbf r}\
({\mathbf F^{\dag}}\times{\mathbf F}-{\mathbf F}\times{\mathbf
F^{\dag}}) \ ,
\end{eqnarray}
where ${\mathbf F^{\dag}}={\mathbf E}-i{\mathbf B}$ is the
hermitian adjoint of ${\mathbf F}$ and the noncommutation of the
fields has been taken in to account. Using the index notation we
have
\begin{eqnarray}
  H&=& \frac{\delta_{kl}}{16\pi}\int\!\!\! d^{3}{\mathbf r}\
  (F^{\dag}_{k} F_{l}+F_{l} F^{\dag}_{k})\ ,\\
  P_{j} &=& \frac{-i\varepsilon_{jkl}}{16\pi c}\int\!\!\! d^{3}{\mathbf r}\
(F^{\dag}_{k} F_{l}+F_{l} F^{\dag}_{k}) \ .
\end{eqnarray}

The same as is the case in classical physics, the hamiltonian and
the momentum are the \emph{generators} of translations in time
and space. This means that the space and time derivatives of a
field operator (an operator depending on $({\mathbf r},t)$) are
given by the commutator with momentum and with the hamiltonian
respectively\cite{wein}. That is
\begin{eqnarray}
  \frac{1}{c}\:\partial_{t}\circ &=& \frac{i}{\hbar c}\left[ H \, ,\,\circ\,\right]\ ,\\
 \partial_{j}\circ &=& \frac{1}{i\hbar}\left[ P_{j}\, ,\,\circ\,\right] \ .
\end{eqnarray}
Where the symbol $\circ$ denotes a space holder to be occupied by
any field operator. Note that even though these two relations are
formally similar, they have completely different origin. The
first relation follows from the time evolution of an operator in
Heisenberg picture whereas the second is a consequence of the
canonical commutation relation between momentum and the position
observable. Using these derivatives in Eqs. (16,17), Maxwell's
equations for the electromagnetic field operators become
\begin{eqnarray}
(\delta_{jk}\delta_{ls}-\delta_{jl}\delta_{sk}+
\delta_{js}\delta_{kl})M_{kls}({\mathbf r},t)&=& 0\ ,\\
  \varepsilon_{kls}M_{kls}({\mathbf r},t)&=& 0 \ ,
\end{eqnarray}
where
\begin{equation}\label{M1}
    M_{kls}({\mathbf r},t)=\int\!\!\! d^{3}{\mathbf r'}
\left[F_{k}^{\dag}({\mathbf r'},t)F_{l}({\mathbf r'},t)+
F_{l}({\mathbf r'},t)F_{k}^{\dag}({\mathbf r'},t) \ ,\
F_{s}({\mathbf r},t)\right]\ .
\end{equation}
In the derivation of this result we have used the important
identity $\varepsilon_{jkl}\varepsilon_{jsu}=
\delta_{ks}\delta_{lu}-\delta_{ku}\delta_{ls}$. We will give a
much simpler expression for $M_{kls}({\mathbf r},t)$ after
obtaining an expression for the fields commutators.
\section{ELECTROMAGNETIC FIELD COMMUTATORS}
We will first obtain an expression for the equal-time field
commutators derived from a quantization \emph{ansatz} and the
requirement of invariance under the ${\mathcal P, T, C}$ and
${\mathcal D}$ transformations. Furthermore we will use the
condition that physical space is homogeneous and isotropic. Once
we know the equal-time commutators we can, in principle,
determine the commutators for different times performing a time
evolution on one of the fields. The transformation behaviour of
the field \emph{operators} are given by
\begin{eqnarray}
  {\mathcal P}\left[ {\mathbf F}({\mathbf r},t)\right]&=&
{-\mathbf F}^{\dag}({-\mathbf r},t)\ ,\\
 {\mathcal T}\left[ {\mathbf F}({\mathbf r},t)\right] &=&
{\mathbf F}({\mathbf r},-t)\ ,\\
{\mathcal C}\left[ {\mathbf F}({\mathbf r},t)\right]&=&
{-\mathbf F}({\mathbf r},t)\ ,\\
{\mathcal D}\left[ {\mathbf F}({\mathbf r},t)\right]&=& {i\mathbf
F}({\mathbf r},t)\ .
\end{eqnarray}
These transformations for the operator fields are essentially the
same as the transformation for the classical fields except for
the time inversion shown in Eqs. (19) and (32). The reason for
this difference is that in quantum mechanics, time inversion is
represented in the Hilbert space by an \emph{anti-unitary}
operator that implies a complex conjugation (the necessity of
this complex conjugation appears clearly upon observation of
Schr{\"o}dinger's or Dirac's equations). Therefore, the imaginary
part of Eq. (15) changes sign twice; once due to the time
inversion property of the magnetic field shown in Eq. (12) and a
second time due to the required complex conjugation.

Let us postulate that the equal-time field commutators, that is,
when both fields are evaluated at the same time $t$ (not shown)
but at different locations, have the form
\begin{eqnarray}
  \left[F_{k}^{\dag}({\mathbf r'})\ ,\ F_{l}({\mathbf r}) \right]&=&
 \alpha_{kl}({\mathbf r'-r})\ ,\\
\left[F_{k}({\mathbf r'})\ ,\ F_{l}({\mathbf r}) \right]&=&
 \beta_{kl}({\mathbf r'-r})\ ,
\end{eqnarray}
where $\alpha_{kl}({\mathbf r'-r})$ and $\beta_{kl}({\mathbf
r'-r})$ are two \emph{functions}, not operators, that we must
determine (of course at the right hand side of these equations an
identity operator is implicit). In this \emph{ansatz} we have
imposed the condition that physical space is homogeneous because
it does'n matter at what space points the commutators are
considered, that is, they can only depend on the separation $
{\mathbf\mbox{\boldmath$\rho$}=r'-r}$. The invariance under
${\mathcal C}$ transformation excludes the possibility that the
commutator results in any odd power of the fields because the
commutator involves the product of two fields and remains
invariant under ${\mathcal C}$; therefore a term linear in the
fields at the right hand side of Eq. (35) and (36) above is
excluded (but not a bilinear form). If we perform the hermitian
conjugation of Eq. (35) (that is, complex conjugation on the
right hand side), with the usual rules for the conjugation of a
product, we reach the conclusion that
\begin{equation}\label{ans}
\alpha_{lk}(-\mbox{\boldmath$\rho$})=
\alpha_{kl}^{\ast}(\mbox{\boldmath$\rho$})\ .
\end{equation}
Hermitian conjugation of Eq. (36) does not results in any
condition for $\beta_{kl}(\mbox{\boldmath$\rho$})$.

Let us first require invariance of the commutation relation in
Eq. (36) under the dual transformation. The right hand side is of
course invariant but the left hand side changes sign. Therefore
we must have
$\beta_{kl}(\mbox{\boldmath$\rho$})=-\beta_{kl}(\mbox{\boldmath$\rho$})$,
that is, $\beta_{kl}(\mbox{\boldmath$\rho$})=0$. This dual
transformation places no constraints on the function
$\alpha_{kl}(\mbox{\boldmath$\rho$})$ that will be fixed by other
symmetry requirements. Let us now impose invariance under time
inversion in Eq.(35) recalling that this transformation implies a
complex conjugation. Eq.(32) indicates that the commutator
remains unchanged and therefore we must have
$\alpha_{kl}(\mbox{\boldmath$\rho$})=\alpha_{kl}^{\ast}(\mbox{\boldmath$\rho$})$,
that is, $\alpha_{kl}(\mbox{\boldmath$\rho$})$ is a \emph{real}
function and with this Eq.(\ref{ans}) becomes
\begin{equation}\label{anst}
\alpha_{kl}(\mbox{\boldmath$\rho$})
=\alpha_{lk}(-\mbox{\boldmath$\rho$}) \ .
\end{equation}
Let us now look what the invariance under ${\mathcal P}$ implies.
Let $\pi$ denote the intrinsic parity of
$\alpha_{kl}(\mbox{\boldmath$\rho$})$ under this transformation,
that is, it may be a tensor or a pseudo tensor (it could even
have mixed parity). Then, using Eq.(31) in the commutator of
Eq.(35) we obtain
\begin{equation}\label{par}
\alpha_{kl}(\mbox{\boldmath$\rho$})
=-\pi\alpha_{lk}(-\mbox{\boldmath$\rho$}) \ ,
\end{equation}
and therefore, comparing with Eq.(38) we conclude that $\pi=-1$,
that is, $\alpha_{kl}(\mbox{\boldmath$\rho$})$ is a pseudotensor.

As a function of $\mbox{\boldmath$\rho$}$,
$\alpha_{kl}(\mbox{\boldmath$\rho$})$ can be decomposed in an
even (\emph{Gerade}) part plus an odd (\emph{Ungerade}) part
under the change
$\mbox{\boldmath$\rho$}\rightarrow-\mbox{\boldmath$\rho$}$,
symbolically $\alpha=\alpha^{G}+\alpha^{U}$, and as a tensor, it
can be decomposed in a symmetric part plus an anti-symmetric part
under the permutation of the indices $kl\rightarrow lk$,
symbolically $\alpha=\alpha^{S}+\alpha^{A}$. Therefore we have
$\alpha=\alpha^{SG}+\alpha^{AG}+\alpha^{SU}+\alpha^{AU}$ but
Eq.(\ref{anst}) implies that the $AG$ and the $SU$ part must
vanish an we are left with only two parts
$\alpha=\alpha^{SG}+\alpha^{AU}$.

Let us first analyse the $AU$ part and we will later see that the
$SG$ part must vanish. It is easy to prove that any
anti-symmetric tensor can be build by the contraction of
$\varepsilon_{kls}$ with an appropriately chosen vector.
Therefore we have
\begin{equation}\label{au}
\alpha_{kl}^{AU}(\mbox{\boldmath$\rho$})
=\varepsilon_{kls}\alpha_{s}(\mbox{\boldmath$\rho$}) \ ,
\end{equation}
where $\alpha_{s}$ is a component of an odd vector:
$\alpha_{s}(\mbox{\boldmath$\rho$})=
-\alpha_{s}(-\mbox{\boldmath$\rho$})$ that we must still
determine (we use the same letter $\alpha$ to denote the tensor
and the vector but this should cause no confusion). Note that
$\varepsilon_{kls}$ is a pseudotensor and therefore
$\alpha_{kl}^{AU}(\mbox{\boldmath$\rho$})$ has the required
parity $\pi=-1$.

Let us now look at the symmetric-even part $\alpha^{SG}$. As a
matrix, it has the form
\begin{displaymath}
\alpha^{SG}(\mbox{\boldmath$\rho$})=\left(
\begin{array}{ccc}
b_{1} & a_{3} & a_{2} \\
a_{3} & b_{2}  & a_{1}  \\
 a_{2} & a_{1} & b_{3}  \\
\end{array}
\right)
\end{displaymath}
where $a_{i}$ and $b_{i}$ are \emph{even} functions, that is,
$a_{i}(-\mbox{\boldmath$\rho$})=a_{i}(\mbox{\boldmath$\rho$})$
and
$b_{i}(-\mbox{\boldmath$\rho$})=b_{i}(\mbox{\boldmath$\rho$})$;
but, if they don't vanish, we have a contradiction with the
requirement that $\alpha_{kl}(\mbox{\boldmath$\rho$})$ is a
pseudo tensor ($\pi=-1$), that is, it must change sign under a
space inversion transformation $\mathcal{P}$. Therefore we must
have $a_{i}=b_{i}=0$. Summarizing, we have found that the
commutation relations of the electromagnetic fields have the form
\begin{eqnarray}
\left[F_{k}({\mathbf r'})\ ,\ F_{l}({\mathbf r}) \right]&=&
\left[F_{k}^{\dag}({\mathbf r'})\ ,\ F_{l}^{\dag}({\mathbf r}) \right]=0 \ ,\\
\left[F_{k}^{\dag}({\mathbf r'})\ ,\ F_{l}({\mathbf r})
\right]&=& \varepsilon_{kls}
 \alpha_{s}({\mathbf r'-r})\ ,
\end{eqnarray}
where $\alpha_{s}(\mbox{\boldmath$\rho$})$ is a component of an
odd vector $\alpha_{s}(\mbox{\boldmath$\rho$})=-
\alpha_{s}(-\mbox{\boldmath$\rho$})$. In order to determine this
vector let us use Eq.(27) and calculate a derivative
$\partial_{s}$ of any field component $F_{l}$ with the condition
that $s\neq l$, that is, $\delta_{sl}=0$.
\begin{eqnarray}
\nonumber \partial_{s}F_{l}({\mathbf r}) &=&
\frac{1}{i\hbar}\left[ P_{s}\,
,\,F_{l}({\mathbf r})\right] \\
\nonumber    &=& \frac{1}{i\hbar} \frac{-i}{16\pi c} \int\!\!\!
d^{3}{\mathbf r'}\ \varepsilon_{skj}\left[F_{k}^{\dag}({\mathbf
r'})F_{j}({\mathbf r'})+ F_{j}({\mathbf r'})F_{k}^{\dag}({\mathbf
r'}) \ ,\
F_{l}({\mathbf r})\right]\\
\nonumber     &=&\frac{-1}{16\pi\hbar c} \int\!\!\! d^{3}{\mathbf
r'}\ \varepsilon_{skj}\ 2\ F_{j}({\mathbf r'})\
\left[F_{k}^{\dag}({\mathbf r'}) \ ,\ F_{l}({\mathbf
r})\right]\\
\nonumber     &=&\frac{-1}{8\pi\hbar c} \int\!\!\! d^{3}{\mathbf
r'}\ \varepsilon_{skj}\varepsilon_{klu}
\ \alpha_{u}({\mathbf r'-r})\ F_{j}({\mathbf r'})   \\
\nonumber   &=&\frac{-1}{8\pi\hbar c} \int\!\!\! d^{3}{\mathbf
r'}\ (-\delta_{sl}\delta_{ju}+\delta_{su}\delta_{jl})
\ \alpha_{u}({\mathbf r'-r})\ F_{j}({\mathbf r'})\\
\nonumber   &=&\frac{-1}{8\pi\hbar c} \int\!\!\! d^{3}{\mathbf
r'}\ \alpha_{s}({\mathbf r'-r})\ F_{l}({\mathbf r'})\ .
\end{eqnarray}
This last line reminds us the definition of the derivative of
Dirac's delta distribution. We have then
\begin{equation}\label{dir}
    \alpha_{s}({\mathbf r'-r})=
8\pi\hbar c\ \partial_{s'}\delta^{3}({\mathbf r'-r})\ ,
\end{equation}
where the primed index $s'$ indicates that we are deriving with
respect to the components of ${\mathbf r'}$. With this we can
finally write the equal time field commutators
\begin{eqnarray}
\left[F_{k}({\mathbf r'},t)\ ,\ F_{l}({\mathbf r},t) \right]&=&
\left[F_{k}^{\dag}({\mathbf r'},t)\ ,\ F_{l}^{\dag}({\mathbf r},t) \right]=0 \ ,\\
\left[F_{k}^{\dag}({\mathbf r'},t)\ ,\ F_{l}({\mathbf r},t)
\right]&=& 8\pi\hbar c\ \varepsilon_{kls'}
 \partial_{s'}\delta^{3}({\mathbf r'-r})\ ,
\end{eqnarray}
that written in terms of the more familiar electric and magnetic
fields are,
\begin{eqnarray}
\left[E_{k}({\mathbf r'},t)\ ,\ E_{l}({\mathbf r},t) \right]&=&
\left[B_{k}({\mathbf r'},t)\ ,\ B_{l}({\mathbf r},t) \right]=0 \ ,\\
\left[E_{k}({\mathbf r'},t)\ ,\ B_{l}({\mathbf r},t) \right]&=&
-i4\pi\hbar c\ \varepsilon_{kls'}
 \partial_{s'}\delta^{3}({\mathbf r'-r})\ .
\end{eqnarray}
Now we will see that the commutation relations found are
compatible with Maxwell's equations. For this, we can first
calculate a simple expression for the tensor $M_{kls}({\mathbf
r},t)$ in Eq.(30) using the commutators given in  Eqs. (44) and
(45).
\begin{eqnarray}
M_{kls}({\mathbf r},t)&=& 16\pi\hbar c\
\varepsilon_{ksu'}\int\!\!\! d^{3}{\mathbf r'}\
 \partial_{u'}\delta^{3}({\mathbf r'-r})\ F_{l}({\mathbf r'},t)\\
&=& -16\pi\hbar c\ \varepsilon_{ksu}\
 \partial_{u}F_{l}({\mathbf r},t)\ .
\end{eqnarray}
Replacing it in Eq. (28) we find that the first Maxwell equation
is identically satisfied whereas the second Maxwell Eq. (29) is
satisfied, provided that $\partial_{u}F_{u}({\mathbf r},t)=0$.
This is a known fact, that the second Maxwell equation in quantum
field theory must be introduced as a subsidiary condition. We can
prove that this subsidiary condition is \emph{necessary} for
logical consistency, showing that otherwise we arrive at an
absurd result:
\begin{eqnarray}
\nonumber \partial_{u}F_{u}({\mathbf r}) &=&
\frac{1}{i\hbar}\left[ P_{u}\,
,\,F_{u}({\mathbf r})\right] \\
\nonumber    &=& \frac{1}{i\hbar}
\frac{-i\varepsilon_{ukl}}{16\pi c} \int\!\!\! d^{3}{\mathbf r'}\
\left[F_{k}^{\dag}({\mathbf r'})F_{l}({\mathbf r'})+
F_{l}({\mathbf r'})F_{k}^{\dag}({\mathbf r'}) \ ,\
F_{u}({\mathbf r})\right]\\
\nonumber     &=&\frac{-\varepsilon_{ukl}}{16\pi\hbar c}
\int\!\!\! d^{3}{\mathbf r'}\ 2F_{l}({\mathbf r'})\
\left[F_{k}^{\dag}({\mathbf r'}) \ ,\ F_{u}({\mathbf
r})\right]\\
\nonumber     &=&\frac{-\varepsilon_{ukl}}{8\pi\hbar c}
\int\!\!\! d^{3}{\mathbf r'}\ F_{l}({\mathbf r'})\ 8\pi\hbar c\
\varepsilon_{kus'}\partial_{s'}\delta^{3}({\mathbf r'-r})\\
\nonumber &=&(\delta_{uu}\delta_{ls'}-\delta_{us'}\delta_{lu})
\int\!\!\! d^{3}{\mathbf r'}\ F_{l}({\mathbf r'})\
\partial_{s'}\delta^{3}({\mathbf r'-r})\\
\nonumber &=&2\delta_{ls'}\int\!\!\! d^{3}{\mathbf r'}\
F_{l}({\mathbf r'})\
\partial_{s'}\delta^{3}({\mathbf r'-r})\\
\nonumber &=&2\ \partial_{l}F_{l}({\mathbf r}) \ .
\end{eqnarray}
One can also show, after some lengthly calculations, that the
commutation relations for the fields given in Eqs. (44) and (45)
lead to the expected commutation relations for the hamiltonian
and momentum, namely $[H\ ,\ P_{k}]=[P_{k}\ ,P_{l}]=0$. In order
to prove this we must use the subsidiary condition
$\partial_{u}F_{u}({\mathbf r},t)=0$ and that the rand terms in
partial integrations vanish because the fields vanish at
infinity.

With the knowledge of the equal time commutators given in Eqs.
(44) and (45), we could calculate the commutators for different
times performing a time evolution in one of the fields with the
hamiltonian given in Eq.(24). For instance we have
\begin{equation}
    \left[F_{k}^{\dag}({\mathbf r'},t')\ ,\ F_{l}({\mathbf r},t)
\right] = \left[\exp\left(\frac{i}{\hbar}(t'-t)H\right)\
F_{k}^{\dag}({\mathbf r'},t)\
\exp\left(\frac{-i}{\hbar}(t'-t)H\right)\ ,\ F_{l}({\mathbf r},t)
\right]\ .
\end{equation}
This would result in
\begin{eqnarray}
\left[F_{k}({\mathbf r'},t')\ ,\ F_{l}({\mathbf r},t) \right]&=&
-8\pi\hbar\ \varepsilon_{kls'}
 \partial_{s'}\partial_{t'}\ D({\mathbf r'-r},t'-t)\ ,\\
\left[F_{k}^{\dag}({\mathbf r'},t')\ , \ F_{l}^{\dag}({\mathbf
r},t) \right]&=& 8\pi\hbar\ \varepsilon_{kls'}
 \partial_{s'}\partial_{t'}\ D({\mathbf r'-r},t'-t)\ ,\\
\left[F_{k}^{\dag}({\mathbf r'},t')\ ,\ F_{l}({\mathbf r},t)
\right]&=& 8\pi\hbar c\ \left(\partial_{k'}\partial_{l'}-
\frac{\delta_{kl}}{c^{2}}\partial_{t'}^{2} \right)D({\mathbf
r'-r},t'-t)\ ,
\end{eqnarray}
where
\begin{equation}\label{D}
D(\mbox{\boldmath$\rho$},\tau) =
\frac{1}{4\pi\rho}\left(\delta(\rho+c\tau)-\delta(\rho-c\tau)\right)
\ ,\ \rho=|\mbox{\boldmath$\rho$}|\ .
\end{equation}
This is not a simple calculation and we will not perform it here.
We just quote the result\cite{com} which, given in terms of the
more familiar electric and magnetic fields, is
\begin{eqnarray}
\left[E_{k}({\mathbf r'},t')\ ,\ E_{l}({\mathbf r},t) \right]&=&
i4\pi\hbar c\
\left(\partial_{k'}\partial_{l'}-\frac{\delta_{kl}}{c^{2}}\partial_{t'}^{2}
 \right) D({\mathbf r'-r},t'-t)\ , \\ \left[B_{k}({\mathbf r'},t')\ ,\ B_{l}({\mathbf
r},t) \right]&=& i4\pi\hbar c\
\left(\partial_{k'}\partial_{l'}-\frac{\delta_{kl}}{c^{2}}\partial_{t'}^{2}
 \right) D({\mathbf r'-r},t'-t) \ ,\\
\left[E_{k}({\mathbf r'},t')\ ,\ B_{l}({\mathbf r},t) \right]&=&
i4\pi\hbar\ \varepsilon_{kls'}
\partial_{s'}\partial_{t'}\ D({\mathbf r'-r},t'-t) \ .
\end{eqnarray}
These commutation relations where first obtained in 1928 by
Jordan and Pauli\cite{jp} but in the context of lagrangian
quantum field theory.

The next step in the program of the quantization of the
electromagnetic field would be to determine the eigenvalues and
eigenvectors for the field operators. However, in the following
section we will see that this is not meaningful and we will
instead define another program consisting in the construction of
the electromagnetic fields as emergent properties of a set of
more fundamental entities.
\section{INTERPRETATION AND CONCLUSION}
In this section we will consider the physical meaning of the
commutation relations of the electromagnetic field. Through the
uncertainty principle, the commutation relations characterize the
compatibility of the corresponding observables. That is, if the
observables commute then there are quantum states where both
observables are precisely determined; otherwise, the product of
their indeterminacies has a lower bound proportional to the value
of the commutator.

The function $D({\mathbf r'-r},t'-t)$ defined in Eq.(54) vanishes
everywhere except at the light cone, that is, except for those
space time points $({\mathbf r'},t')$ and $({\mathbf r},t)$ that
can be connected by a light ray. Therefore the electromagnetic
field for pairs of points that can not be joined by a light ray
can assume exact values simultaneously because their commutator
vanish. For points on the light cone, the commutator is highly
singular; it involves second order derivatives of the singular
distribution $D({\mathbf r'-r},t'-t)$.

If these non vanishing commutators would take some \emph{finite}
value we would have a situation not unusual in quantum mechanics
where the indeterminacy principle limits the simultaneous
measurability of two observables but at least one of them could
be known exactly and in all other cases both observables could be
determined within some finite range of indeterminacy. Furthermore
in the limit $\hbar\rightarrow 0$ we would recover the classical
behaviour.

In the case of the electromagnetic fields, the \emph{singular}
character of the commutators have unacceptable consequences:
assume that we could know the fields at one point $({\mathbf
r},t)$; then they would propagate, even a short distance, to
$({\mathbf r'},t')$ where they would have an infinite
indeterminacy. In this sense, the quantum mechanical treatment of
the electromagnetic fields does not appear to be very reasonable.
L. D. Landau and R. Peierls\cite{land} in 1931 recognized this
difficulty and, without reference to the commutators but with
some heuristic (but correct) arguments, reached the conclusion
that for quantum theory ``the field strength are not measurable
quantities''. In an answer to this important remark, published
two years later, N. Bohr and L. Rosenfeld\cite{bohr} noted that
the commutation relations among \emph{field averages}, taken on a
finite volume, have non singular and finite values (depending on
the size and shapes of the volumes where the averages are taken)
and since every laboratory measurement of the fields involve an
extended test charge, implying that what we really measure are
field averages and not the fields at a point, they concluded that
the difficulty can be dismissed because it has no consequence in
a laboratory measurement. Rosenfeld commented\cite{ros} that for
them the ``field components taken at a definite space time point
are used in the formalism as an idealization without immediate
physical meaning; the only meaningful statements of the theory
concern averages of such field components over finite space time
regions''.

The typically positivistic solution given by Bohr and Rosenfeld
to the problem posed by the singular commutation relations of the
fields in not satisfactory for those of us that believe that
physics is not \emph{only} concerned with real laboratory
measurements but it also deals with idealizations and models that
we build in order to describe reality. For this group there are
two possibilities open: either is quantum mechanics wrong when
applied to the electromagnetic field or the electromagnetic field
is an essentially macroscopic system such that a quantum
treatment of it is meaningless. Of course the enormous success of
quantum mechanics in all realms of physics makes the second
possibility more plausible. There are  several examples of
emergent properties of physical systems that do not make any
sense at a deep microscopic level and whose quantization would be
meaningless. For instance a temperature field $T({\mathbf r},t)$,
or pressure or specific volume of a gas correspond to collective
observables of large number of molecules. The molecules can, or
must, be treated with quantum mechanics but to quantize pressure
or temperature is meaningless. Already at the classical level
there is an indication that the electromagnetic field is not a
basic or elementary system. This comes from the fact that the
energy of the system does not scales with size of the system.
Consider a physical system  with fields $\mathbf{E_{1}}$ and
$\mathbf{B_{1}}$ with an energy $H_{1}$ and another system with
fields $\mathbf{E_{2}}$ and $\mathbf{B_{2}}$ with an energy
$H_{2}$. Now, the \emph{compound} system with fields
$\mathbf{E=E_{1}+E_{2}}$ and $\mathbf{B=B_{1}+B_{2}}$ will
\emph{not} have the energy $H_{1}+H_{2}$ as one would expect from
the combination of two elementary systems. The question is then,
what are the corresponding ``molecules'' of the electromagnetic
fields? The answer is well known: the photons.

Consequently we postulate that the electromagnetic field is a
collective emergent properties of an ensemble of more fundamental
entities, the photons, carriers of energy, momentum and angular
momentum. The quantum mechanical description of the photons and
how this set of bosons build the electromagnetic field will be
the subject of following works.

It is relevant to mention that the choice adopted here, that the
photons are the \emph{primary ontology} that must be treated
quantum mechanically and that the electromagnetic fields are
macroscopic manifestations of an ensemble of them, is not a
unique choice. There are indeed authors that consider the
electromagnetic fields as the primary ontology and that the
photons do not have an objective existence but are rather
mathematical entities, ``particle-like excitations''\cite{bal},
corresponding to the normal mode decomposition of the fields.
This alternative option can be adopted because, although the
objective existence of the photons provide the most natural
explanation of the photoelectric effect and of Compton
scattering, these are not compelling evidence for their existence
because these effects can also be explained by a semiclassical
theory combining unquantized fields with quantum theory of
matter\cite{semiclas1,semiclas2}. The existence of these two
alternative interpretations indicate that the quantum description
of electromagnetic radiation is not a closed subject and that
Einstein's question \emph{What are light quanta?} has not been
answered.
\begin{acknowledgements}
I would like to thank H. M{\'a}rtin for challenging discussions and
E. Fern{\'a}ndez for bibliographic assistance. This work received
partial support from ``Consejo Nacional de Investigaciones
Cient\'{\i}ficas y T\'ecnicas'' (CONICET).
\end{acknowledgements}


\begin{thebibliography}{99}
\bibitem{units} The Heaviside system is the ``right''
one where the electric and magnetic fields have the same units,
as they should because they are different components of the same
tensor, and the $4\pi$ is in Coulomb law, where it belongs
because of its inherent spherical symmetry. Unfortunately, the SI
system of units has hit the wrong choice.
\end{thebibliography}
\end{document}